# Conceptual Design of Pumping Systems up to 1000 $m^3$/s Based on Non-Evaporable Getters in a Gas-Dynamic Magnetic Trap

Alexey M. Semenov[1,2], Nataliy G. Vasilieva[1], Sergey V. Polosatkin[1,2,3]
[1]Budker Institute of Nuclear Physics, Novosibirsk, Russia
[2]Novosibirsk State Technical University, Novosibirsk, Russia
[3]Novosibirsk State University, Novosibirsk, Russia

Corresponding author: A.M.Semenov@inp.nsk.su

*Abstract*— The Gas-Dynamic Magnetic Trap (GDMT), a new linear magnetic facility for confining high-temperature hydrogen plasma, is currently under development at the Budker Institute of Nuclear Physics (BINP SB RAS). The total hydrogen flow that must be removed from the end expanders by the vacuum system is approximately $10^{22}$ molecules/s, which corresponds to a required pumping speed on the order of 500 $m^3$/s. Non-evaporable getters (NEG) are considered as a possible pumping method for this application. This article presents experimental studies of Ti–Zr–Al getters manufactured by Polema LLC. The pumping speed, sorption capacity, and sticking factors of the getter surface were measured. The sticking factors for hydrogen and deuterium are 2.23% and 1.54%, respectively. The getter pumping speed is maintained over 60 air-vent/reactivation cycles. Based on these results, a conceptual design of a NEG-based pumping system for the GDMT end expanders is proposed, providing a specific pumping speed of 74 $m^3/(m^2 \cdot s)$ for the getter panel. The proposed pumping system design and the required heating power are discussed.

## I. Introduction

Linear magnetic systems for confining hot plasma are promising configurations for creating a nuclear fusion reactor. Research on plasma confinement in linear systems, conducted at the Budker Institute of Nuclear Physics (BINP), has made it possible to formulate a concept for a fusion reactor based on a linear magnetic configuration [1-4]. To experimentally verify the confinement principles underlying the reactor designs, a new experimental magnetic trap with gas-dynamic plasma confinement, the GDMT, is currently being developed at the BINP[5]. An important feature of this facility, which distinguishes it from existing devices, is its long pulse duration (up to 100 seconds), significantly exceeding the plasma confinement time in the trap. The long pulse duration makes it possible to study steady-state (in terms of particle balance) operating regimes of the facility, but at the same time imposes demanding requirements on its engineering systems.

The GDMT facility in its basic configuration consists of a central section, magnetic mirror plug units, and end expanders (Fig. 1). Hydrogen plasma, heated in the central section to thermonuclear temperatures, flows out through the magnetic plugs along the magnetic field lines and strikes the plasma dumps located in the end expanders, where it is neutralized and converted into a flow of molecular hydrogen. This flow must be removed from the expanders by a vacuum pumping system. At the same time, secondary ionization of hydrogen in the expander plasma leads to an increase in the plasma density, a change in the electric potential distribution along the magnetic field lines, and, consequently, a reduction in the plasma confinement efficiency of the facility. Experiments performed on the GDT facility [6] have shown that, in order to maintain efficient plasma confinement, the neutral gas density in the expander must not exceed $10^{13}$cm$^{-3}$ (0.04 Pa).

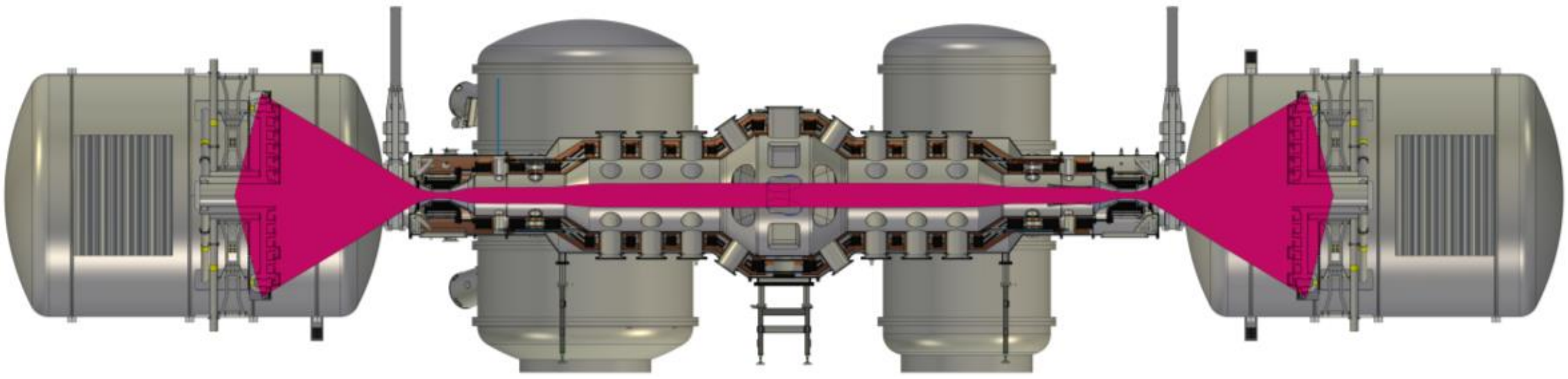

Fig.1. GDMT facility layout.

The expected plasma flux from the central section of the GDMT into the expander is up to $10^{22}$ s$^{-1}$, which, when recalculated as the flux of molecular hydrogen desorbed from the surface of plasma dumps, corresponds to 18.5 m$^3$·Pa/s. Thus, the expander pumping system must provide a pumping speed on the order of 500 m$^3$/s throughout the facility's operational pulse (up to 100 s). Note that such a pumping speed corresponds to a perfectly absorbing surface area of 1.1 m$^2$, meaning that the required pumping can only be achieved by large-area pumping modules with high hydrogen capture efficiency.

Pumping of large hydrogen throughputs is a common issue in fusion research. This challenge is particularly acute in the design of neutral beam injectors, where a pressure gradient from about 10 Pa in the ion source and neutralizer to ~$10^{-4}$ Pa in the beam transport section must be sustained. The vacuum systems of neutral beam injectors feature pumping speeds in excess of 1000 m$^3$/s [7-9]. For instance, the pumping system developed for the ITER injectors provides a pumping speed of 5000 m$^3$/s at a throughput of 34 m$^3$·Pa/s [10].

As a baseline option for the expander pumping system, a system based on cryogenic modules located in the end tank behind the plasma dump is being considered. In this case, the pumping speed of this system is limited by the vacuum conductance of the gap between the plasma dump and the tank wall. In the present paper, we propose a pumping system based on non-evaporable getters (NEG) located directly in front of a surface of the plasma pumps, which can be used either in combination with the cryogenic pumping system or serve as the primary means of pumping the expanders.

## II. Non-evaporatible getters pumping

Non-evaporable getters (NEG) are materials consisting of alloys of aluminum, zirconium, vanadium, titanium, and rare-earth metals. These materials feature a high solubility for gases and high diffusion coefficients which makes them suitable for vacuum pumping of all gases except noble gases and methane. The high ultimate solubility of oxygen for elements of group IV of the periodic table allows for multiple passivation (opening to the atmosphere)/reactivation cycles at the activation temperature without a noticeable decrease in the pumping speed. NEG alloys were well studied for the possibility of application in ultra-high vacuum [11]. For the first time, they were used in ultrahigh vacuum system as well as in the Large Electron Positron collider yet in 1970s [12]. And since that time NEG has been an integral part of vacuum technology [13,14]. In synchrotron light machines NEG are used to minimize the gas load [15] as distributed coatings. The next step in vacuum technology was the development of combined NEG and sputter ion pumps [16] able to evacuate inert gases and methane for the high vacuum and ultra-high vacuum.

Among active gases, interacting with the getter, hydrogen and its isotopes have a peculiar behavior. They are sorbed in a reversible way. In typical conditions, hydrogen dissociates at the solid surface, and diffuses in the getter material. This provides a stable pumping speed over time as the bulk of the material makes a contribution to the sorption capacity, not only the surface. A limit to the sorption capacity for NEG elements is cracking due to saturation by hydrogen, when subject to cyclic hydrogen loads reaching concentrations above the recommended ones. The behavior under sorption and regeneration conditions allows the NEG application in machines requiring large hydrogen pumping. Experimental prototypes of NEG-based pumps were tested at QUEST and EAST tokamaks [17-18]. Conceptual design of scalable vacuum pump for large hydrogen throughput systems like neutral beam injectors based on sintered ceramics ZAO was developed [19].

## III. Description of NEG-Based Vacuum Pump Prototype

A prototype of NEG-based vacuum pump, originally developed at the BINP for use in vacuum system of “SKIF“ synchrotron radiation source [20-22], was subsequently studied as a candidate for pumping large hydrogen gas loads. The prototype is based on Ti–Zr–Al discs produced by Polema LLC [23]. Such getter discs are produced in large quantities for the nuclear power and oil and gas industries.

The main parameters of the prototype are summarized in Table I. The prototype consists of getter discs stacked in six columns, with 31 discs per column. Each disc has an outer diameter of 25 mm, an inner diameter of 8 mm (for accommodating the heater), and a thickness of 1.6 mm. A porosity of the disk material is of 30%. The discs are arranged with a 1.5 mm gap between adjacent discs. The overall height of the pump, measured from the flange to the end face, does not exceed 130 mm. Temperature is monitored using a chromel–alumel thermocouple. A protective screen is installed on the outside to reduce the heater power by half. The pump is equipped with Conflat flanged connections. Photographs of the pump are shown in Fig. 2.

TABLE I. Main Parameters of the NEG Prototype

| Name | Activation Temperature, °C | NEG Surface, $cm^2$ | NEG Weight, g | Porosity, % | Flange Type |
|---|---|---|---|---|---|
| NNG-1.3 | 650 | 1873 | 360 | 30 | DN 100 |

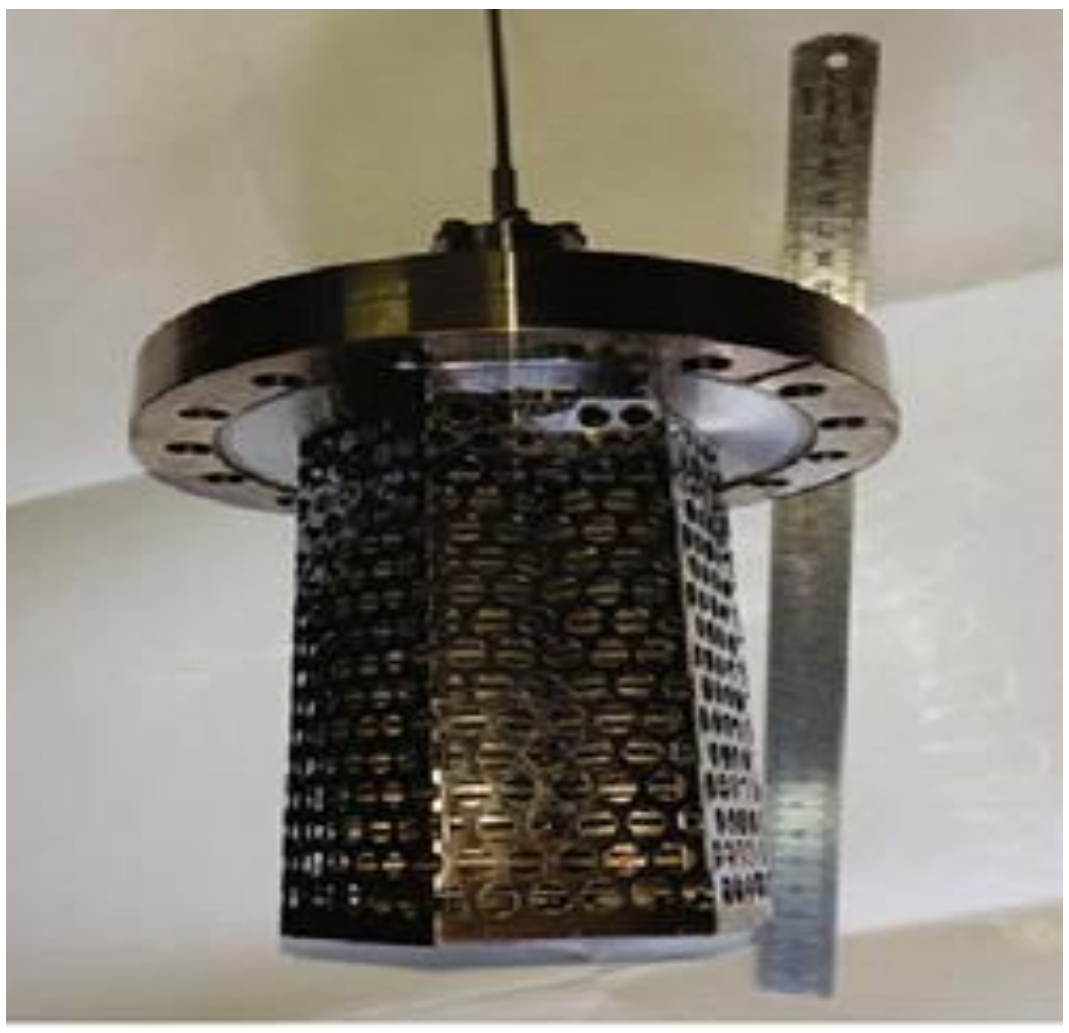

Fig. 2. Images of getter pump NNG-1.3.

## IV. Experimental Setup and Measuring Technique

The activation and gas sorption characteristics of the getter pump were measured on a specialized experimental setup (Fig. 3). The system was preliminarily evacuated by a station with a turbomolecular pump (*TMP1*) and an oil-free diaphragm pump (*MP*). High-vacuum pumping was achieved by the turbomolecular pump *TMP2* through vacuum all-metal valves *VR2*. The typical pressure level of turbomolecular pump *TMP2* is equal down to $10^{-8}$ Pa. A mechanical pressure gauge and a Baratron capacitive gauge were used to monitor the pressure of the injected gas.

Gas was let into the system through leak valve *VF* and capillary *C* (the molecular conductivity with respect to hydrogen was 2.8 $10^{-3}$ l/s). The injection system was evacuated by the turbomolecular pump *TMP1* and an oil-free diaphragm pump *MP* through vacuum all-metal valves *VR4* and *VR5*. Emergency electric valve *VR3* was installed to prevent the atmosphere from breaking through from the *TMP2* and *TMP1*. The fore and high vacuums were measured at the pumping station by a Pfieffer PKR 251 full range pressure sensor consisting of a Pirani sensor and a cold cathode gauge.

The pressure in the system was measured by hot cathode gauge *IG1*. The partial pressures of gases were measured using an *RGA* quadrupole mass spectrometer from SRS (Stanford, United States). Hot cathode gauge *IG2* was needed to control the gas flow pumped out by turbomolecular pump *TMP2*.

It was measured the effective pumping speed of turbomolecular pump *TMP2*, the molecular conductivity of the capillary, and the coefficients of sensitivity of the pressure sensors with hot cathodes for each gas ($H_2$, CO and $D_2$). Table II gives these values for each gas. All vacuum gauges were calibrated each time after vent of air atmosphere. Valve *VR1* was closed at the start of the experiments. It was then opened, and the injected gas began to flow into the test chamber through the capillary.

The gas flow into the test chamber can be determined from:

$$Q_{total} = C \cdot (P_{bar} - P_{IG1}), \tag{1}$$

where $C$ is the conductance of the calibrated capillary; $P_{bar}$ and $P_{IG1}$ are the pressures measured by the capacitance gauge and the hot-cathode gauge.

Pumping speed, $S_{NEG}$, of the getter during gas injection is given by:

$$S_{NEG} = \frac{C \cdot (P_{bar} - P_{IG1})}{K \cdot \Delta P_{IG1}} - S_{TMP2}, \qquad (2)$$

where $S_{NEG}$ is the pumping speed of the NEG pump. $S_{TMP2}$ is the pumping speed of the turbomolecular pump (28 l/s for hydrogen, 8.5 l/s for carbon monoxide and 24 l/s for deuterium); $C$ is the molecular conductivity of the capillary; $\Delta P_{IG1}$ is the pressure rise on hot cathode gauge *IG1* after opening of VR1 valve; $K$ is the coefficient of sensitivity of the gauge to the injected gas. Note that the pressure-jump method used here measures the pumping speed governed by gas capture (i.e., the sticking factor), and is therefore not affected by back-flow from the getter due to saturation.

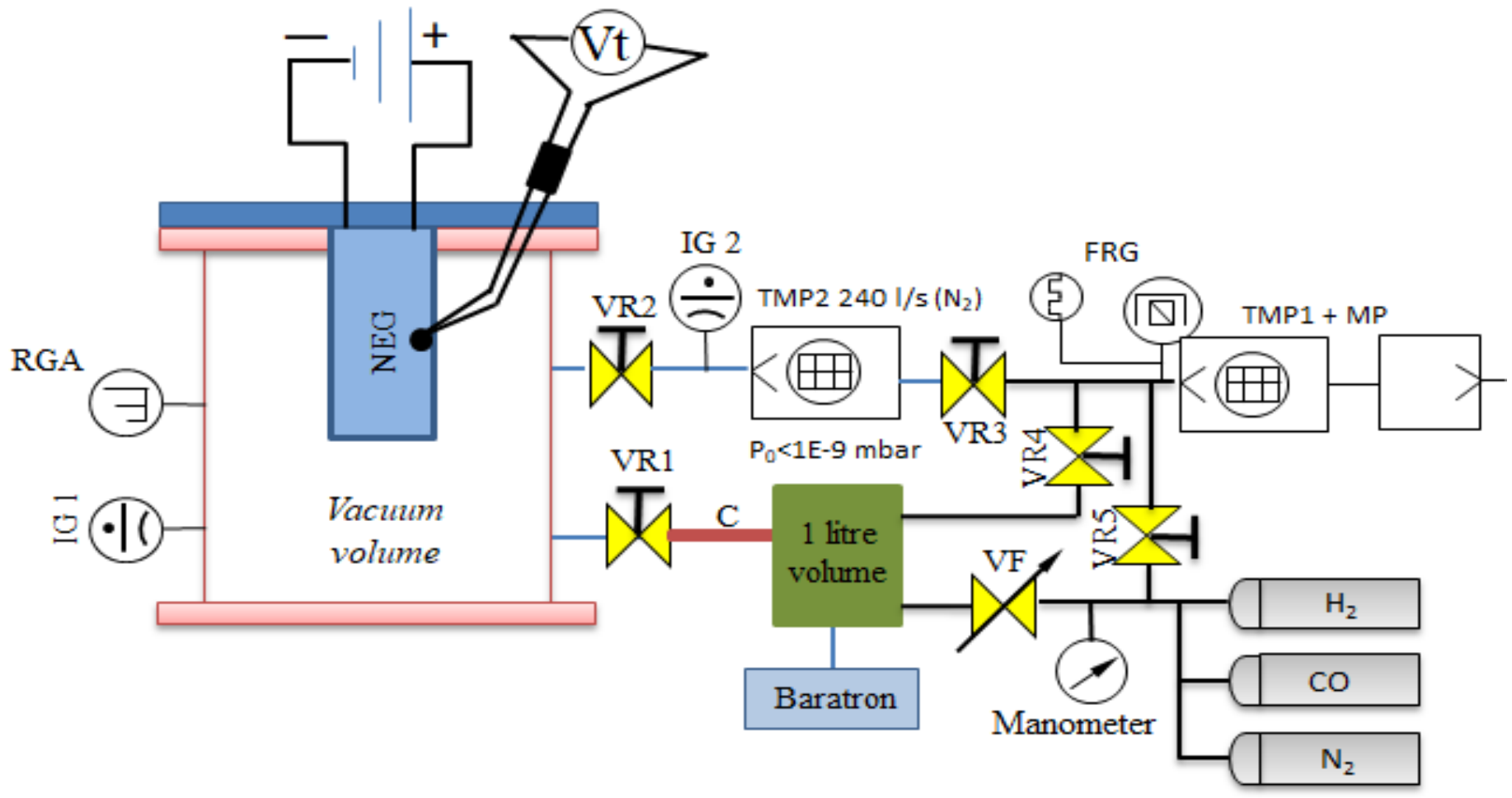

Fig. 3. Simplified scheme of the vacuum setup. RGA is a mass spectrometer; IG1 and IG2 is a hot-cathode gauges; FRG is a full-range sensor; NEG is the pump being tested; C is a capillary with known molecular conductance; TMP1+MP is a turbomolecular pumping station; TMP2 is a turbomolecular pump; the Baratron is a capacitive gauge; the manometer is mechanical pressure gauge; VF is a slotted leak; VR1, VR2, VR4, and VR5 are all-metal angle valves; and VR3 is the emergency electric valve.

TABLE II. MOLECULAR CONDUCTIVITY OF THE CAPILLARY, EFFECTIVE PUMPING SPEED OF TURBOMOLECULAR PUMP 2, AND COEFFICIENT OF SENSITIVITY OF HOT CATHODE GAUGES FOR DIFFERENT GASES

| Gas | Molecular conductance C, l/s | Pumping speed of $S_{TMP2}$, l/s | Coefficient of sensitivity, $K_{IG1}$ | Coefficient of sensitivity, $K_{IG2}$ |
|---|---|---|---|---|
| $H_2$ | 2,8E-3 | 28 | 2,3 | 2,3 |
| $D_2$ | 2,15E-3 | 21 | 3,55 | 3,55 |

The quantity of gas sorbed by the getter was determined as the difference between the amount of gas introduced and the amount pumped out by the turbomolecular pump.

$$q = \int_0^t (Q_{total} - Q_{TMP2}) \cdot dt, \quad (3)$$

where $Q_{total}$ is the gas flow rate introduced into the chamber; $Q_{TMP}2$ is the flow rate pumped out by the turbomolecular pump *TMP2*; and *t* is the integration time.

The gas flow rate pumped out by the turbomolecular pump is determined as:

$$Q_{TMP2} = S_{TMP2} \cdot K \cdot (\Delta P_{IG1} - \Delta P_{IG2}), \quad (4)$$

where $\Delta P_{IG1}$, $\Delta P_{IG2}$ are the pressure rise on hot cathode gauges IG1 and IG2 after opening of VR1 valve, and *K* is the coefficient of sensitivity of the gauges to the injected gas.

## V. Results and Discussions

After pumping down and leakage check at room temperature, each time before measurement the getter was activated for 1 h at the maximum temperature (650°C), and passivated in air for at least 3 h after each measurement. The quantity of activations and passivations was equal 60 cycles. A larger number of cycles is expected.

Fig. 4 shows the pumping speed of the NNG-1.3 pump for hydrogen and deuterium as a function of the sorbed gas quantity. The pumping speeds of the fresh getter are found to be 1.3 $m^3/s$ for hydrogen and 0.7 $m^3/s$ for deuterium. The pumping speed remains constant for both hydrogen and deuterium up to a sorbed hydrogen quantity of 2 $Pa \cdot m^3/g$. Multiple activation/passivation cycles do not affect the hydrogen pumping capability. A further increase in the sorbed quantity leads to a slow decrease in pumping speed ($S \sim q^{-0.2}$). At the ultimate sorbed quantity of 1230 $Pa \cdot m^3/g$, which is close to the getter embrittlement limit, the pumping speed decreases by a factor of 3.

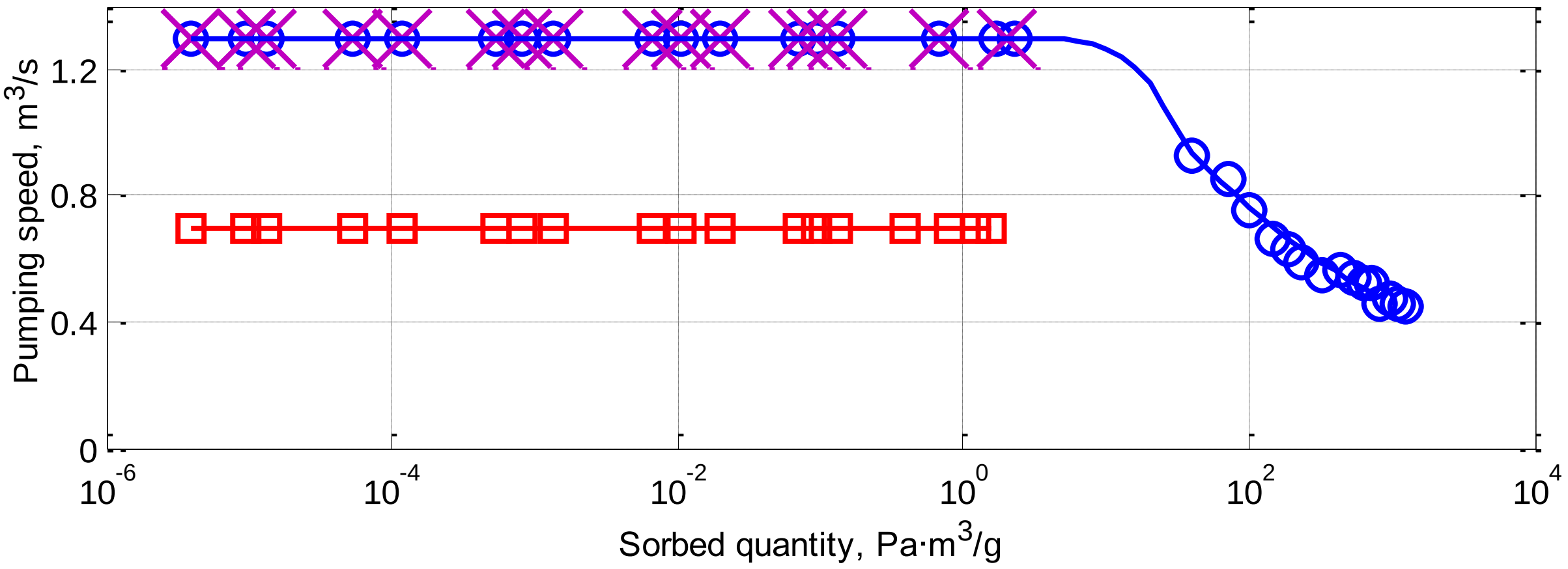

Fig.4. The NNG-1.3 pumping speed for hydrogen and deuterium as a function of the sorbed gas quantity. Rings and squares: hydrogen and deuterium, fresh pump. Crosses: hydrogen, after 60 activation/passivation cycles.

In principle, the chemical activity of hydrogen and its isotopes does not depend on the isotope mass. Under such conditions, the normalized pumping speed and sticking factors of hydrogen and deuterium should be equal. However, the diffusion coefficient of a dissociated atom from the surface into a bulk material depends on its mass, and this effect can affect the overall pumping speed [24]. According to [24], another explanation may be that even a small content of helium impurities (less than 1%) can lead to a decrease in the pumping speed by a factor of two for deuterium. However, any of these statements requires additional experimental research or the creation of a sound physical model.

To determine the sticking factor of the getter used, gas propagation in the NNG-1.3 pump was simulated using Molflow+ code [25]. The figure 5 shows the calculated dependence of the NNG-1.3 pumping speed on the getter sticking factor. The points correspond to the measured pumping speed values for hydrogen and deuterium. The resulting sticking factor values are 2.23% and 1.54%, respectively. The presence of the thermal shield and self-shielding by the getter disks reduces the pump efficiency by 30% compared to pumping by a free getter surface. The pressure distribution for hydrogen (sticking factor 2.23%) across the pump cross-section is shown in Fig. 6.

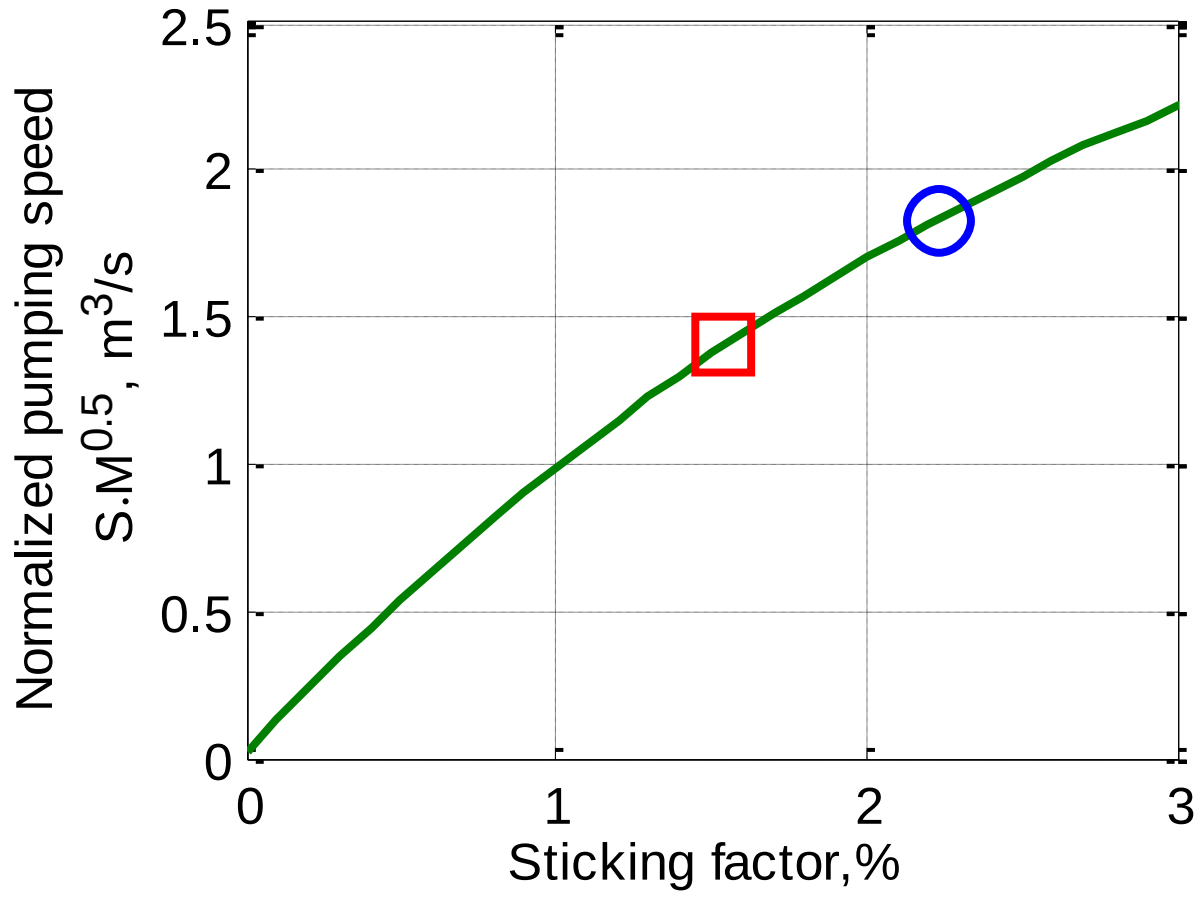

Fig. 5. Calculated dependence of NNG-1.3 pumping speed on the getter sticking factors. Dots correspond the measured pumping speed for hydrogen (ring) and deuterium (square)

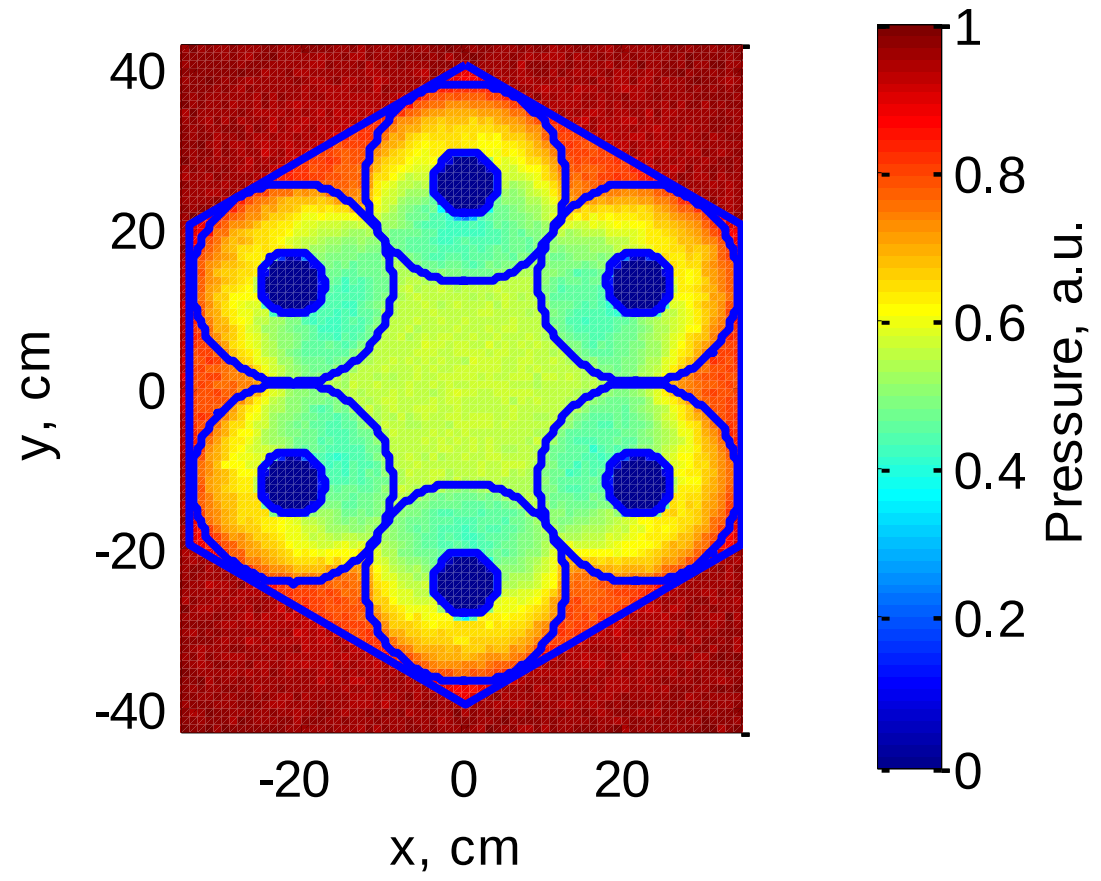

Fig. 6. Calculated pressure distribution across NNG-1.3 cross-section for getter sticking factor 2.23%

The equilibrium constants in Sieverts' law were determined for the Ti–Zr–Al getter used in this work. For this purpose, the equilibrium pressure was measured as a function of hydrogen content and temperature over the ranges of 150–650 °C (in 50 °C steps) and 0.1–4 Pa·m³/g, respectively. In total, 150 points were measured at different temperatures and concentrations. The dependence of the equilibrium pressure on hydrogen content can be expressed as:

$$\lg\left(P_{[\mathrm{Pa}]}\right) = (7.44 \pm 0.48) + 2 \cdot \lg\left(q_{\left[\frac{\mathrm{Pa}\cdot\mathrm{m}^3}{\mathrm{g}}\right]}\right) - \frac{5300 \pm 280}{T_{[\mathrm{K}]}} \qquad (5)$$

Fig. 7 shows this dependence together with the experimentally measured points for several temperatures.

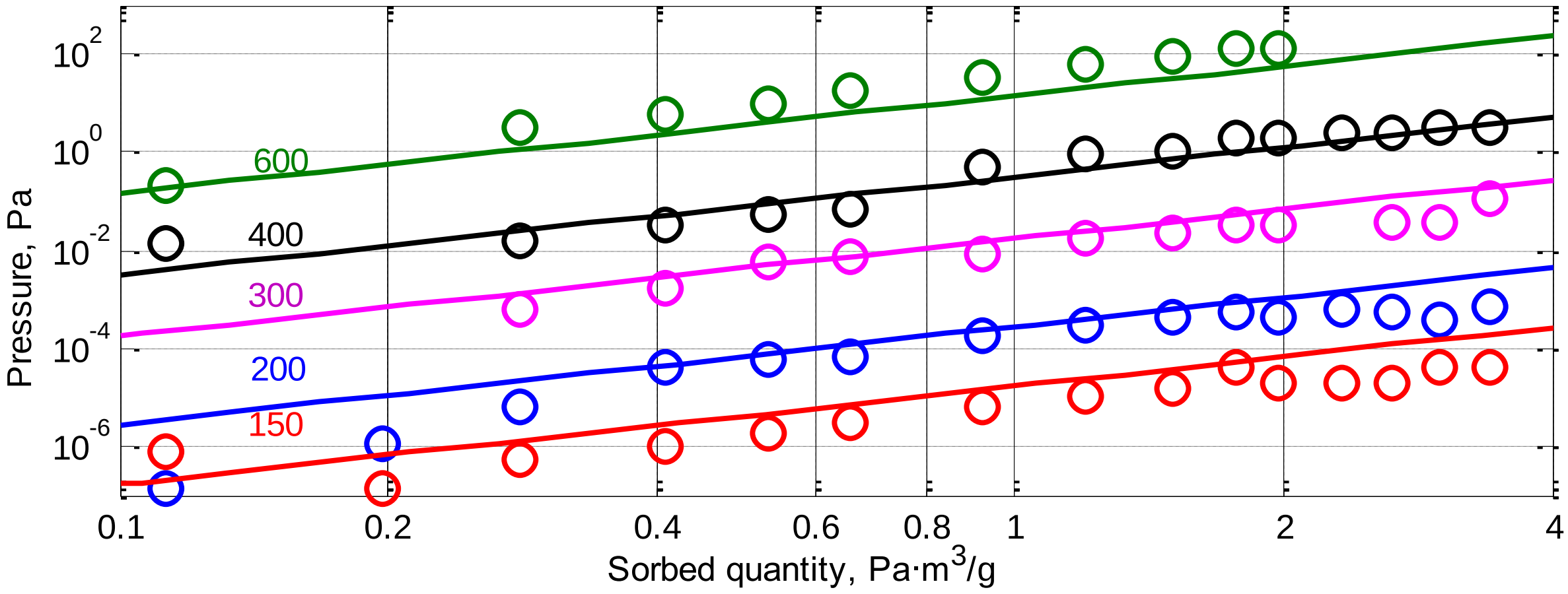

Fig. 7. Hydrogen equilibrium pressure versus sorbed hydrogen quantity. Symbols – experimental data, lines – fits by equation (5). Numbers near the lines indicate getter temperature in °C.

## VI. NEG-Based Vacuum Pump for GDMT end expanders

Although the required pumping capacity of the GDMT end expander pumping system is considerably higher than the pumping speed of the NNG-1.3 pump used as a prototype, the system can be constructed from the same basic components: columns of NEG discs with an internal heater. The length, number, and relative positioning of these columns are dictated by the design and the required pumping speed. Since the sticking factor of the NEG disc surface is rather low, the disc arrangement must be optimized to reduce full length of the columns. Modeling of the pumping performance of a disc column was carried out, with the gap between adjacent discs being the variable parameter.

To estimate the required column length, we will use the parameter "effective sticking factor" — the probability of a hydrogen molecule being absorbed by the outer surface of the column. The dependence of the effective sticking factor on the gap between the discs is shown in Fig. 8a. This dependence has a maximum, which is related to the fact that a small gap causes strong self-shielding of the disk surfaces, while a large gap reduces the absorber area per unit column length. The maximum effective sticking factor, equal to 9.2% and corresponding to a pumping speed of 3.2 m³/s per meter of column, is achieved at a gap of 0.8 mm. For a gap of 1.5 mm, used in the NNG-1.3 pump, the effective sticking factor decreases to 8.6% (3.0 m³/s per meter of column), but the efficiency of getter utilization increases (i.e., the number of discs required to provide the required pumping speed). This efficiency can be characterized by the mean (over the getter surface) sticking factor. The dependence of the mean sticking factor on the gap between the discs is shown in Fig. 8b. It is equal to 1.57% for a gap of 0.8 mm and 1.88% for a gap of 1.5 mm, i.e., increasing the gap allows the number of getters required to provide pumping at the required speed to be reduced by 20%.

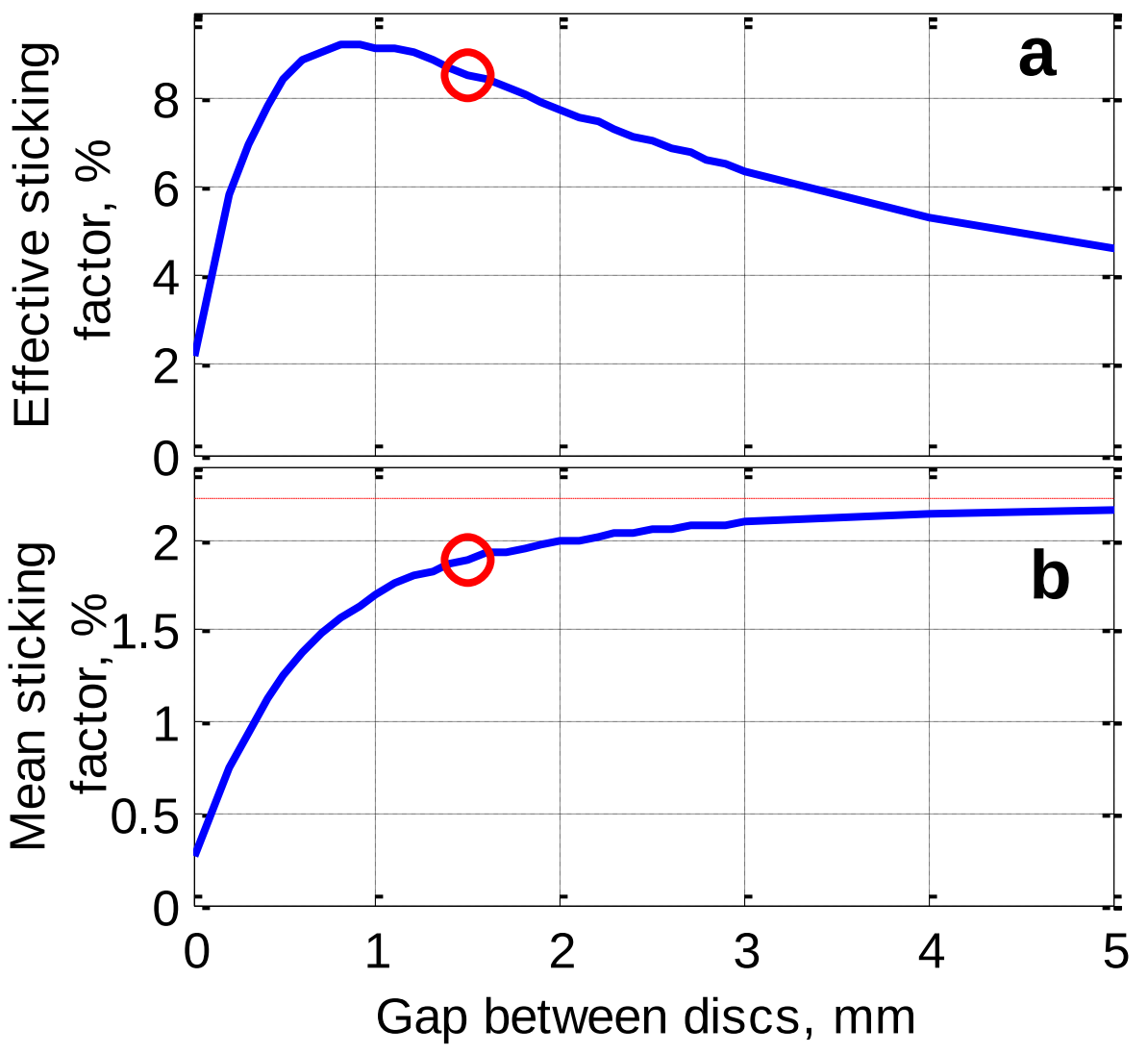

Fig. 8. Dependence of effective sticking of getter column surface (a) and mean sticking factor over getter surface (b) from the distance between discs. Circles shows the values for the NNG-1.3 pump. Dashed line correspond to sticking of free getter surface (2.23%)

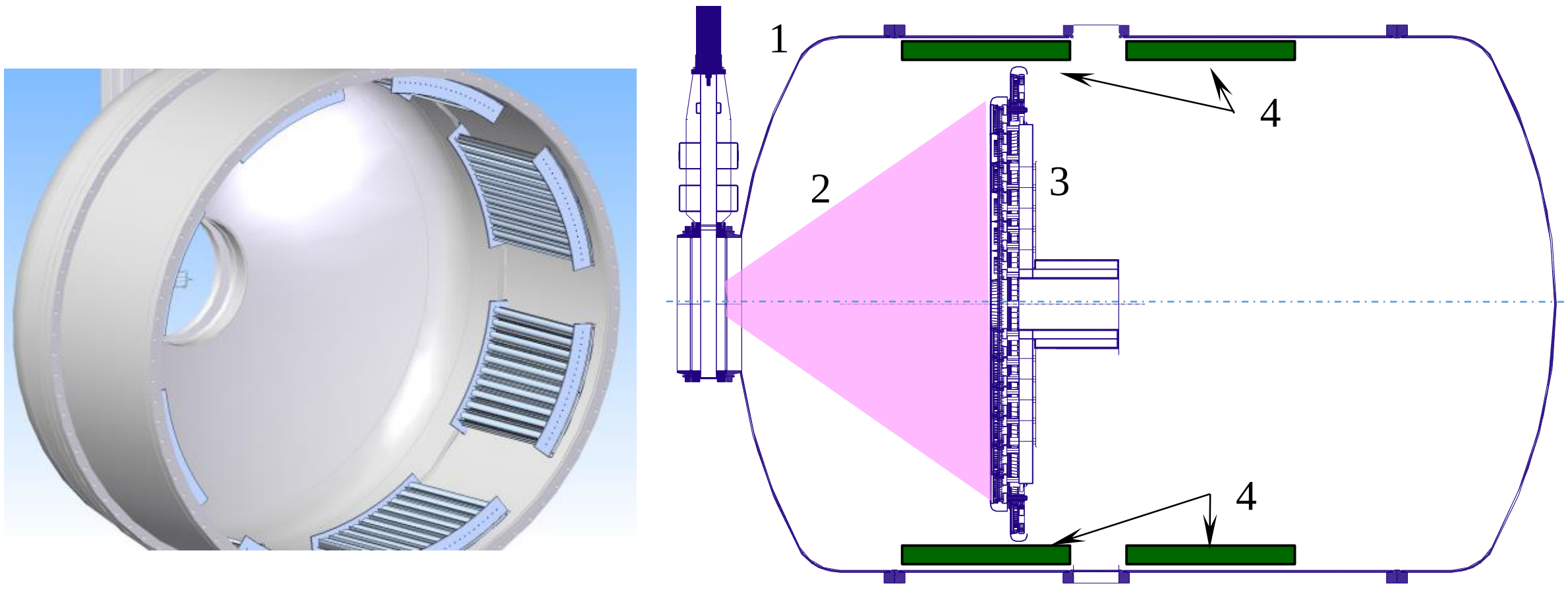

Fig. 9. End expander. 1 – vacuum chamber, 2 – plasma stream, 3 – the plasma receiver, 4 – getter columns

The end expander design is shown in Fig. 9. It consists of a cylindrical vacuum volume 4 m in diameter with a flat plasma receiver located inside. The most convenient and technologically feasible arrangement is to place the getter columns parallel to the axis of the facility, close to the cylindrical wall of the vacuum chamber. The getter columns are combined into panels equipped with thermal shields, which reduce the heat flux to the chamber wall during getter activation, and with current leads for powering the heaters.

Since in this design the pumped gas can enter only from one side of the panel, calculations of the panel pumping speed as a function of the distance between the columns in the panel were performed.

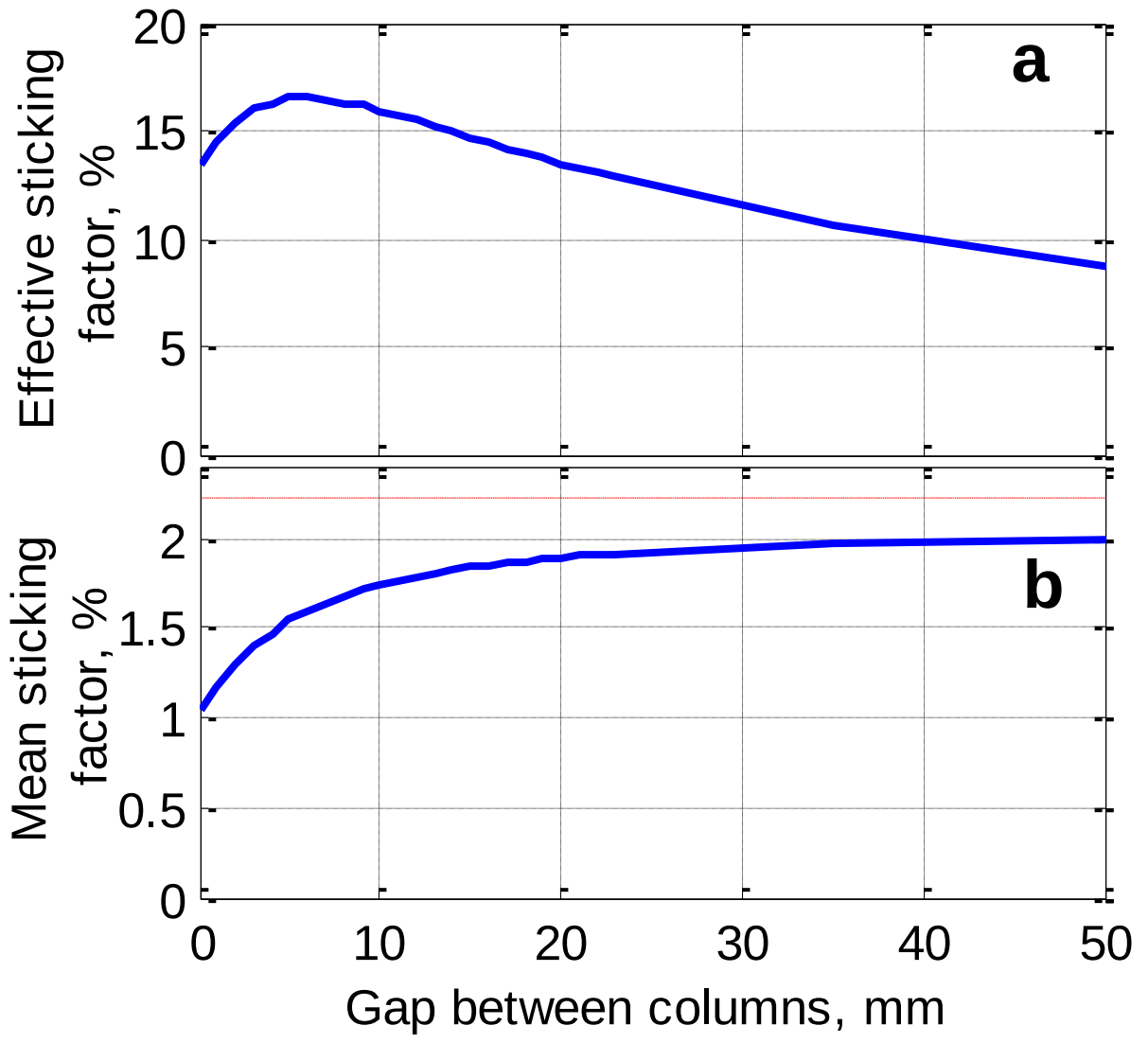

Fig. 10. Dependence of effective sticking of the pumping panel (a) and mean sticking factor over getter surface (b) from the distance between columns in the panel. Dashed line correspond to sticking of free getter surface (2.23%)

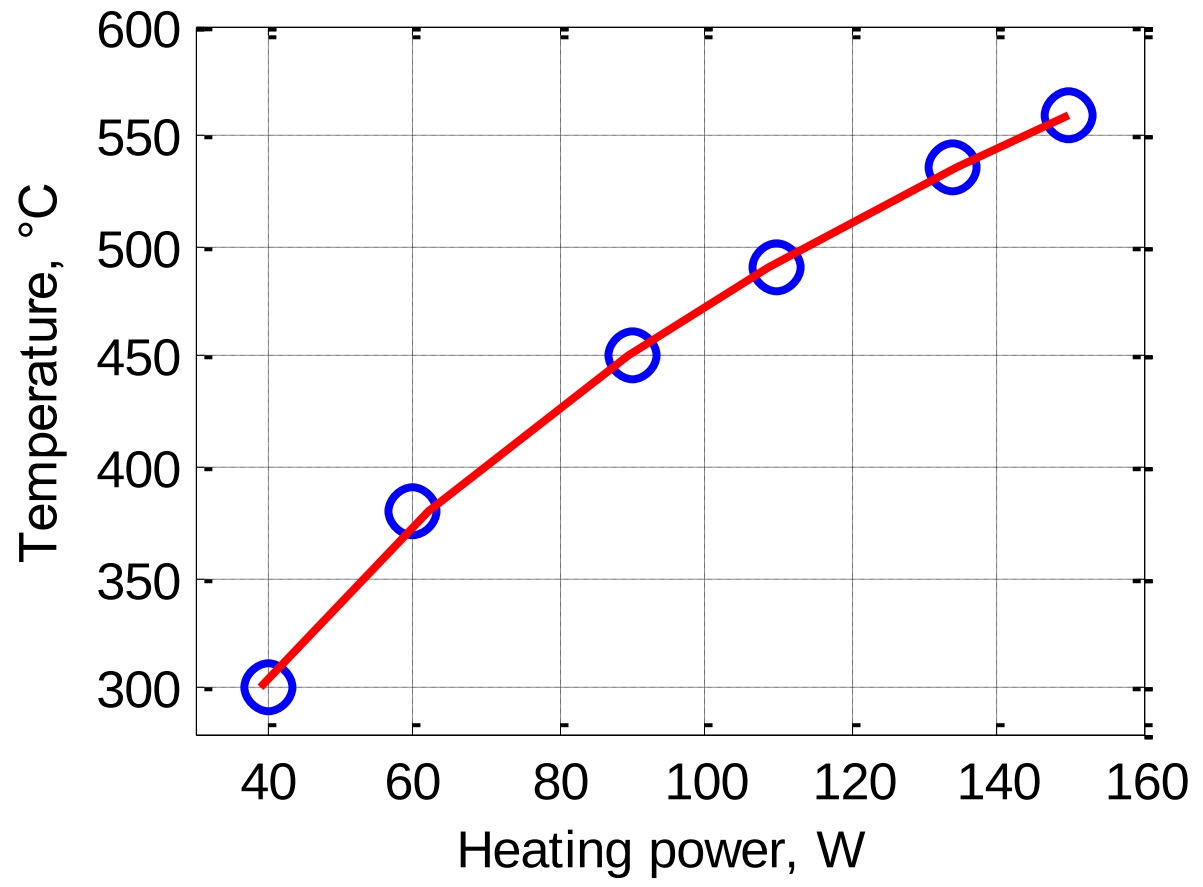

Fig. 11. Getter column temperature vs. heating power. Circles correspond to measurements, line – fit with emissivity coefficient 0.64.

Results of these calculation are shown in Fig.10. Maximal effective sticking factor for the panels achieves 16.5% for distance between columns of 6 mm, that means the panels should cover 6.8 $m^2$ of surface of the vacuum chamber to provide pumping speed 500 $m^3/s$ (specific pumping speed of 74 $m^3/(m^2 \cdot s)$ per unit panel area). The required getter mass is 140 kg. This mass is sufficient to ensure 20,000 seconds of facility operation (note that the explosion hazard of hydrogen limits the total duration of facility operation without hydrogen removal to 14,000 s).

For the NEG pumps the power required for getter activation is important issue. The NEG temperature vs. heating power was measured for single NEG column without thermal shield in vacuum (fig.11). Based on this measurement we evaluate effective emissivity coefficient for the NEG column of 0.64. Using this emissivity, we performed radiative heat transfer simulations in COMSOL Multiphysics. The results show that the radiated power at the getter activation temperature is 590 W per meter of column length. For the entire pumping system with a capacity of 500 $m^3$/s, heating to activation temperature therefore requires a thermal power of 130 kW.

## VII. Conclusions

Non- evaporable getters fabricated by Polema LLC were tested for possibility to capture reversibly huge amount of hydrogen. It is found that the getters are suitable for use in the high-throughput hydrogen pumping systems for fusion facilities. In particular the pump system based on NEG can provide 500 $m^3$/s pumping for end expander of GDMT facility.

## VIII. Acknowledgements

The work is supported by Russian Science Foundation, grant No. 25-22-00117